\documentclass[aps,pre,showpacs,superscriptaddress,preprint,letterpaper]{revtex4-1}
\usepackage{amsmath,amssymb,amsfonts,epsfig,graphicx,txfonts,array}

\newcommand*{\dmatul}[2]{\underline{D}_{#1}(#2)}

\newcommand*{\dmatf}[2]{\widetilde{D}_{#1}(#2)}

\newcommand*{\gmatul}[2]{\underline{G}_{#1}(#2)}

\newcommand*{\gmatf}[2]{\widetilde{G}_{#1}(#2)}

\newcommand*{\Rv}[0]{\vec{R}}

\newcommand*{\Rvuab}[2]{\Rv+\vec{x}_{#1}-\vec{x}_{#2}}
\newcommand*{\kv}[0]{\vec{k}}
\newcommand*{\ku}[0]{\hat{k}}
\newcommand*{\kmax}[0]{k_\text{max}}
\newcommand*{\Lmax}[0]{L_\text{max}}
\newcommand*{\Mmax}[0]{M_\text{max}}
\newcommand*{\eye}[0]{\mathbf{1}}
\newcommand*{\DAA}{\widetilde{D}_\mathrm{AA}}
\newcommand*{\DAO}{\widetilde{D}_\mathrm{AO}}
\newcommand*{\DOA}{\widetilde{D}_\mathrm{OA}}
\newcommand*{\DOO}{\widetilde{D}_\mathrm{OO}}
\newcommand*{\GAA}{\widetilde{G}_\mathrm{AA}}
\newcommand*{\GAO}{\widetilde{G}_\mathrm{AO}}
\newcommand*{\GOA}{\widetilde{G}_\mathrm{OA}}
\newcommand*{\GOO}{\widetilde{G}_\mathrm{OO}}

\newlength{\wholefigwidth}
\newlength{\smallfigwidth}
\newlength{\halfsmallfigwidth}
\begin{document}
\title{Direct calculation of lattice Green function with arbitrary interactions for general crystals}
\date{\today}
\author{Joseph A. Yasi}
\email{yasi@illinois.edu}
\affiliation{Department of Physics, University of Illinois at Urbana-Champaign,
Urbana, IL 61801}
\author{Dallas R. Trinkle}
\affiliation{Department of Materials Science and Engineering, University of
Illinois at Urbana-Champaign, Urbana, IL 61801}

\begin{abstract}
Efficient computation of lattice defect geometries such as point defects, dislocations, disconnections, grain boundaries, interfaces and free surfaces requires accurate coupling of displacements near the defect to the long-range elastic strain.  Flexible boundary condition methods embedded a defect in infinite harmonic bulk through the lattice Green function.  We demonstrate an efficient and accurate calculation of the lattice Green function from the force-constant matrix for general crystals with an arbitrary basis by extending a method for Bravais lattices.  New terms appear due to the presence of optical modes and the possible loss of inversion symmetry.  By separately treating poles and discontinuities in reciprocal space, numerical accuracy is controlled at all distances.  We compute the lattice Green function for a two-dimensional model with broken symmetry to elucidate the role of different coupling terms.  The algorithm is generally applicable in two and three dimensions, to crystals with arbitrary number of atoms in the unit cell, symmetry, and interactions.
\end{abstract}


\maketitle

\section{Introduction}
Atomic-scale modeling of lattice defect geometries such as point defects, dislocations, disconnections, grain boundaries, interfaces and free surfaces are key for understanding many material properties\cite{Haasen96}.  The anisotropic elasticity solutions for the displacement fields are accurate at distances far from the defects\cite{Stroh1962,Bacon1979}; however, the solutions frequently diverge near the defect center, requiring an atomistic approach to determine the defect core geometry.  For many defects, especially dislocations, the long-range strain field is incompatible with periodic boundary conditions.  To efficiently model these defects, Sinclair 
\textit{et al.} developed a flexible boundary condition method based on the lattice Green function (LGF) \cite{Sinclair1978}.  The technique was extended for crack propagation\cite{Thomson1992,Canel1995} using empirical potentials and recently applied to isolated dislocations with density functional theory (DFT)\cite{Woodward2002} and empirical potentials\cite{Rao1998,Yang2001}.  Prior to this method, dislocation calculations in DFT were limited to dipole\cite{Arias1994,Frederiksen2003} and quadrupole\cite{Bigger1992,Arias2000} geometries due to the long-range strain field of an isolated dislocation.  The flexible boundary condition method couples the simulation cell to infinite bulk by treating an intermediate region away from the defect core as harmonic and relaxing these forces with a LGF.  Efficient numerical calculation of the LGF for point defects in cubic lattices is well known\cite{Tewary1973,MacGillivray1983,Tewary2004}.  An automated technique for efficiently calculating the lattice Green function with arbitrary atomic interactions was only applicable to Bravais lattices\cite{TrinkleLGF2008}, making it unsuitable for many materials with more than one atom in the crystal basis such as HCP (e.g. Mg and Ti).

We extend this numerical technique to general crystals with arbitrary numbers of atoms in the crystal basis. The extension to multiple atom unit cells requires a separate treatment for acoustic and optical modes in the long wavelength limit, but continues to rely on the same input information (force constants) and shows similar efficiency\cite{Ghazisaeidi2009}.  Section \ref{sec:harmonic} describes harmonic response in a multiatom basis and the general symmetries of the dynamical matrix and lattice Green function.  Section \ref{sec:LGF} describes the numerical technique for efficiently calculating the LGF for general crystals.  Section \ref{sec:square} computes the lattice Green function for a simple doubled square lattice and new terms due to symmetry breaking.  This symmetry breaking is manifest in both reciprocal and real space.  The result is an efficient, automated algorithm for calculating the lattice Green function for general crystals which can efficiently relax defect geometries; in particular, relaxation of dislocation geometries in Mg\cite{Yasi2009},  Ti\cite{Ghazisaeidi2012}, and as part of an interfacial LGF calculation of a twin boundary in Ti\cite{Ghazisaeidi2010}.

\section{Harmonic Response}
\label{sec:harmonic}
Flexible boundary condition techniques allow for efficient calculation of isolated lattice defects with a small number of atoms by using the perfect lattice Green function (LGF) to couple the defect to bulk.  We extend the numerical method for calculating the LGF\cite{TrinkleLGF2008} to work with general crystals with multiple atoms in the unit cell.  The infinite harmonic crystal is well known from classical and quantum theory\cite{BornHuang1954,Maradudin1971}.  For a crystal with $N$ atoms in the basis, the $3N \times 3N$ force-constant matrix $\dmatul{i\alpha,j\beta}{\Rv-\Rv'}$ determines the force on basis atom $i$ at lattice site $\Rv$ in Cartesian direction $\alpha$ from a displacement of a basis atom $j$ at lattice site $\Rv'$ in Cartesian direction $\beta$
\begin{equation}
\dmatul{i\alpha,j\beta}{\Rv-\Rv'} = \frac{\partial^2 U^\text{total}}{\partial u_{i\alpha}(\Rv)\partial u_{j\beta}(\Rv')}\biggr|_{\vec{u}=0}.
\end{equation}
Due to independence of differentiation order and the inversion symmetry of all Bravais lattices, the force-constant matrix obeys $\dmatul{i\alpha,j\beta}{\Rv} = \dmatul{j\beta,i\alpha}{-\Rv}$.  Unlike a Bravais lattice, a general crystal does not necessarily have inversion symmetry.  Therefore, $\dmatul{j\beta,i\alpha}{-\Rv} = \dmatul{j\beta,i\alpha}{\Rv}$ is {\em not} guaranteed.  The force-constant matrix obeys a sum rule,
\begin{equation}
\sum_{\Rv,j}\dmatul{i\alpha,j\beta}{\Rv} = 0,\quad \forall i,\alpha,\beta
\label{eqn:sumrule}
\end{equation}
due the absence of forces under a uniform translation of the crystal.  Under a uniform strain,
there is no net force on the unit cell.  Hence, the force-constant matrix obeys
\begin{equation}
\sum_{\Rv,i,j,\beta}\dmatul{i\alpha,j\beta}{\Rv}(\Rvuab{i}{j})_\beta = 0,\quad \forall \alpha
\end{equation}
where the $\vec{x}_i$ are the positions of the atoms within the unit cell.  Since there is no torque under a uniform rotation, the force-constant matrix also obeys
\begin{equation}
\sum_{\Rv,j}\left(\dmatul{i\alpha,j\beta}{\Rv}(\Rvuab{i}{j})_\gamma - \dmatul{i\alpha,j\gamma}{\Rv}(\Rvuab{i}{j})_\beta\right) = 0, \quad \forall i,\alpha,\beta,\gamma.
\end{equation}
In the harmonic limit, the LGF $\gmatul{i\alpha,j\beta}{\Rv-\Rv'}$ gives displacements $(u_{i\alpha})$ in response to forces $(f_{j\beta})$,
\begin{equation}
u_{i\alpha}(\Rv) = -\sum_{\Rv',j,\beta}\gmatul{i\alpha,j\beta}{\Rv-\Rv'}f_{j\beta}(\Rv')
\end{equation}
where $\alpha$ and $\beta$ index Cartesian directions, $i$ and $j$ index atoms
in the crystal basis, and $\Rv$ and $\Rv'$ are lattice vectors.
The LGF is the pseudoinverse of the force-constant matrix,
\begin{equation}
\sum_{\Rv'',l,\gamma}\dmatul{i\alpha,l\gamma}{\Rv-\Rv''}\gmatul{l\gamma,j\beta}{\Rv''-\Rv'}
= \delta_{ij}\delta_{\alpha\beta}\delta(\Rv-\Rv'),\quad \forall i,j,\alpha,\beta,\Rv,\Rv'.
\label{eqn:lgfinversereal}
\end{equation}
The sum rule (Eq.~\eqref{eqn:sumrule}) guarantees that $\dmatul{i\alpha,j\beta}{\Rv}$ has three zero modes
(uniform translation), and therefore $\dmatul{i\alpha,j\beta}{\Rv}$ is singular.

\section{Computation of the LGF}
\label{sec:LGF}
For computational efficiency and control of numerical errors, we compute the LGF by inverting the dynamical matrix in reciprocal space.  First, we Fourier transform the force-constant matrix to the dynamical matrix.  We then invert the dynamical matrix using a block partitioning scheme by separating the dynamical matrix into acoustic and optical modes in order to isolate the poles and discontinuities.  The inverse contains a first- and second-order poles and a discontinuity in reciprocal space.  For numerical efficiency and stability, we perform the inverse Fourier transform to real space analytically for the poles and discontinuities, and numerically for the semi-continuum correction.  Finally, to get the real space LGF, we rotate back into the crystal coordinate system to complete the calculation.

Computing the LGF is more tractable in reciprocal space, where the Fourier transforms of the LGF (similarly for the force constant matrix and dynamical matrix) are,
\begin{equation}
\gmatf{i\alpha,j\beta}{\kv} = \sum_{\Rv} e^{i\kv\cdot(\Rvuab{i}{j})}\gmatul{i\alpha,j\beta}{\Rv}
,\quad
\gmatul{i\alpha,j\beta}{\Rv} = V\iiint_\textrm{BZ}\displaylimits\negthickspace
\frac{d^3k}{(2\pi)^3}\;e^{-i\kv\cdot(\Rvuab{i}{j})}\gmatf{i\alpha,j\beta}{\kv},
\end{equation}
for unit cell volume $V$.  Note that we choose the Fourier phase factor to correspond to the crystal vector between two atoms.  We use tildes and underlines to represent matrices in reciprocal and real space, respectively.  In reciprocal space, Eq.~\eqref{eqn:lgfinversereal} becomes
\begin{equation}
\sum_{l,\gamma}\gmatf{i\alpha,l\gamma}{\kv}\dmatf{l\gamma,j\beta}{\kv} = \delta_{ij}\delta_{\alpha\beta}\delta(\kv).
\label{eqn:fgmatdmat}
\end{equation}
The sum rule (Eq.~\eqref{eqn:sumrule}) means that for $\kv=0$, $\dmatf{i\alpha,j\beta}{0}$ has three zero eigenvalues corresponding to the three uniform translation modes.  In addition, for small $\kv$, three modes of $\dmatf{i\alpha,j\beta}{\kv}$ will go as $k^2$; this creates a singularity in $\gmatf{i\alpha,j\beta}{\kv}$ at $k=0$, corresponding to both second- and first-order poles.  Hence, we will expand the lattice Green function around $\kv=0$, and solve for the individual terms in the expansion from the expansion of the dynamical matrix around $\kv=0$.

To isolate the second-order pole in the elastic Green function due to translational symmetry, we choose a basis for atomic displacements/forces in the unit cell to separate the acoustic and optical modes at $k=0$.  This involves the eigenvectors of $\dmatf{i\alpha,j\beta}{k=0}$,
\begin{equation}
\sum_{j,\beta}\dmatf{i\alpha,j\beta}{0}e^{\mu}_{i\alpha} = \lambda^{\mu}e^{\mu}_{j\beta};
\end{equation}
we identify the first three $\mu=1\ldots3$ eigenvalues $\lambda^{\mu}=0$ as the acoustic modes, and the remaining $3N-3$ as optical modes with positive eigenvalues.  The acoustic eigenvectors are $e^{\mu}_{j\beta} = \delta_{\mu\beta}/\!\sqrt{N}$ for $\mu = 1 \ldots 3$, and the full set of eigenvectors provide an orthonormal basis.  In this new basis, the dynamical matrix is
\begin{equation}
\dmatf{\sigma\sigma',\mu\nu}{\kv} = \sum_{\Rv,j,\gamma,l,\lambda} \left(1 + (i\kv\cdot(\Rvuab{j}{l})) -
\frac{(\kv\cdot(\Rvuab{j}{l}))^2}{2!} + \cdots\right)
e^{\mu}_{j\gamma}\dmatul{j\gamma,l\lambda}{\Rv}e^{\nu}_{l\lambda}
\label{eqn:drotsum}
\end{equation}
where $\sigma=\text{A}$ for $\mu=1\ldots3$, and $\sigma=\text{O}$ otherwise, with a similar relation between $\sigma'$ and $\nu$.  For the acoustic-acoustic projection ($\sigma=\sigma'=\text{A}$), the zeroth order term is zero due to the sum rule Eq.~\eqref{eqn:sumrule},
\begin{equation}
\frac{1}{N}\sum_{\Rv,j,\gamma,l,\lambda}\delta_{\mu\gamma}\dmatul{j\gamma,l\lambda}{\Rv}\delta_{\nu\lambda} = \frac{1}{N}\sum_{l}\sum_{\Rv,j}\dmatul{j\mu,l\nu}{\Rv} = 0.
\end{equation}
The remaining odd order contributions in $\kv$ to the acoustic-acoustic quadrant are zero due to inversion symmetry of a Bravais lattice ($\Rv \rightarrow -\Rv$).  Thus, the AA term expands as
\begin{equation}
\dmatf{\mathrm{AA},\mu\nu}{\kv} = \sum_{\Rv,j,l} \left(-\frac{(\kv\cdot(\Rvuab{j}{l}))^2}{2!} + \frac{(\kv\cdot(\Rvuab{j}{l}))^4}{4!} + \cdots\right)\dmatul{j\mu,l\nu}{\Rv}.
\end{equation}
Since the leading order term of the acoustic subspace of the dynamical matrix is second order in $k$, the acoustic-acoustic (AA) projection of the LGF will have a second-order pole in $k$.  Similarly, the sum rule also requires the zeroth order term of the acoustic-optical (AO) and optical-acoustic (OA) projections of the dynamical matrix to be zero. The AO and OA projections have first-order poles, and the optical-optical (OO) projection does not have a pole.  In the rotated basis, we can perform a block inversion of the dynamical matrix
\begin{equation}\label{eqn:grotblock}
\begin{split}
\widetilde{G}^\text{ROT} &= \left(
\begin{array}{c|c}
\GAA & \GAO\\
\hline
\GOA & \GOO\\
\end{array}
\right) = \left(
\begin{array}{c|c}
\DAA & \DAO\\
\hline
\DAO^{\dagger}  & \DOO\\
\end{array}
\right)^{-1}\\
&= \left(
\begin{array}{c|c}
\left(\DAA - \DAO\DOO^{-1}\DAO^{\dagger}\right)^{-1} & -\left(\DAA - \DAO\DOO^{-1}\DAO^{\dagger}\right)^{-1}\DAO\DOO^{-1}\\
\hline
-\DOO^{-1}\DAO^{\dagger}\left(\DAA - \DAO\DOO^{-1}\DAO^{\dagger}\right)^{-1}  & \left(\DOO - \DAO^{\dagger}\DAA^{-1}\DAO\right)^{-1}\\
\end{array}
\right)
\end{split}
\end{equation}
where the roman indexes $\mathrm{A}$ and $\mathrm{O}$ correspond to the projection onto the acoustic and optical bases respectively.  This ensures that the $k^{-2}$ divergence as $k \rightarrow 0$ is contained in the acoustic-acoustic quadrant of the matrix.  Divergences of order $ik^{-1}$ can also appear at leading order in the acoustic-optical and optical-acoustic quadrants for crystals without inversion symmetry.  For crystals with inversion symmetry, only the even order terms in the acoustic-acoustic quadrant and the odd order terms in the acoustic-optical and optical-acoustic quadrants of the dynamical matrix remain (Eq.~\eqref{eqn:drotsum}).  We write our expansion of the dynamical matrix and LGF in power series,
\begin{equation}
 \widetilde{D}_{\sigma\sigma',\mu\nu}(\vec{k}) = \sum_{n=0}^{\infty}\widetilde{D}_{\sigma\sigma',\mu\nu}^{(n)}(\ku)(i k)^{n},
\quad
\widetilde{G}_{\sigma\sigma',\mu\nu}(\vec{k}) = \sum_{n=-2}^{\infty}\widetilde{G}_{\sigma\sigma',\mu\nu}^{(n)}(\ku)(i k)^{n},
\end{equation}
where $\sigma$ and $\sigma'$ can be basis $\mathrm{A}$ or $\mathrm{O}$, and $\widetilde{D}_{\sigma\sigma'}^{(n)}$ and $\widetilde{G}_{\sigma\sigma'}^{(n)}$ are the $n$th coefficients of the power series.  We compute the power series coefficients for the dynamical matrix by calculating the $n^\text{th}$ term of the expansion in Eq.~\eqref{eqn:drotsum},
\begin{equation}
\widetilde{D}_{\sigma\sigma',\mu\nu}^{(n)}(\ku) = \sum_{\Rv,j,\gamma,l,\lambda} \frac{(\ku\cdot(\Rvuab{j}{l}))^n}{n!}e^{\mu}_{j\gamma}\dmatul{j\gamma,l\lambda}{\Rv}e^{\nu}_{l\lambda}.
\label{eqn:dmatpowercoeff}
\end{equation}
As stated previously, $\DAA^{(0)} = \DAA^{(1)} = 0, \DAO^{(0)} = \DOA^{(0)} = 0, \GOO^{(-2)} = \GOO^{(-1)} = 0,$ and $\GAO^{(-2)} = \GOA^{(-2)} = 0$.

In terms of these power series coefficients, the divergent and discontinuous terms of the LGF are
\begin{equation}
\begin{split}
 \GAA^{(-2)} &= \Lambda^{(2)-1}\\
 \GAA^{(-1)} &= \Lambda^{(2)-1}\Lambda^{(3)}\Lambda^{(2)-1}\\
 \GAA^{(0)} &= \Lambda^{(2)-1}\Lambda^{(4)}\Lambda^{(2)-1} - \Lambda^{(2)-1}\Lambda^{(3)}\Lambda^{(2)-1}\Lambda^{(3)}\Lambda^{(2)-1}\\
 \GOA^{(-1)} = \GAO^{(-1)\dagger} &= -\DOO^{(0)-1}\DAO^{(1)\dagger}\Lambda^{(2)-1}\\
 \GOA^{(0)} = \GAO^{(0)\dagger} &= \DOO^{(0)-1}\DAO^{(2)\dagger}\Lambda^{(2)-1} - \DOO^{(0)-1}\DOO^{(1)}\DOO^{(0)-1}\DAO^{(1)\dagger}\Lambda^{(2)-1} + \DOO^{(0)-1}\DAO^{(1)\dagger}\Lambda^{(2)-1}\Lambda^{(3)}\Lambda^{(2)-1}\\
 \GOO^{(0)} &= \left(\DOO^{(0)} - \DAO^{(1)\dagger}\DAA^{(2)-1}\DAO^{(1)}\right)^{-1}
\end{split}
\label{eqn:gpower}
\end{equation}
where,
\begin{equation}
\begin{split}
 \Lambda^{(2)} &= -\DAA^{(2)} + \DAO^{(1)}\DOO^{(0)-1}\DAO^{(1)\dagger}\\
 \Lambda^{(3)} &= -\DAO^{(1)}\DOO^{(0)-1}\DAO^{(2)\dagger} - \DAO^{(2)}\DOO^{(0)-1}\DAO^{(1)\dagger} + \DAO^{(1)}\DOO^{(0)-1}\DOO^{(1)}\DOO^{(0)-1}\DAO^{(1)\dagger}\\
 \Lambda^{(4)} &= -\DAO^{(1)}\left(\DOO^{(0)-1}\DOO^{(2)}\DOO^{(0)-1}-\DOO^{(0)-1}\DOO^{(1)}\DOO^{(0)-1}\DOO^{(1)}\DOO^{(0)-1}\right)\DAO^{(1)\dagger}\\
               &\;\;\;\; + \DAO^{(2)}\DOO^{(0)-1}\DAO^{(2)\dagger} + \DAO^{(1)}\DOO^{(0)-1}\DAO^{(3)\dagger} + \DAO^{(3)}\DOO^{(0)-1}\DAO^{(1)\dagger}\\
               &\;\;\;\; - \DAO^{(2)}\DOO^{(0)-1}\DOO^{(1)}\DOO^{(0)-1}\DAO^{(1)\dagger} - \DAO^{(1)}\DOO^{(0)-1}\DOO^{(1)}\DOO^{(0)-1}\DAO^{(2)\dagger} - \DAA^{(4)}
\end{split}
\label{eqn:glambdas}
\end{equation}
For crystals with inversion symmetry, the relations for some of the terms are considerably simplified
\begin{equation}
\begin{split}
 \GOA^{(0)} = \GAO^{(0)\dagger} &= \DOO^{(0)-1}\DAO^{(1)\dagger}\Lambda^{(2)-1}\Lambda^{(3)}\Lambda^{(2)-1}\\
 \Lambda^{(3)} &= \DAO^{(1)}\DOO^{(0)-1}\DOO^{(1)}\DOO^{(0)-1}\DAO^{(1)\dagger}\\
 \Lambda^{(4)} &= \DAO^{(1)}\DOO^{(0)-1}\DAO^{(3)\dagger} + \DAO^{(3)}\DOO^{(0)-1}\DAO^{(1)\dagger} - \DAA^{(4)}
\end{split}
\label{eqn:glambdaszero}
\end{equation}

Inverse Fourier transforming the divergent terms converges slowly.  To efficiently calculate the LGF in real space, we integrate these divergent terms analytically, and integrate the remaining continuous terms numerically.  To facilitate analytic integration, we expand the divergent terms in spherical harmonics for a 3D LGF and in a Fourier series for a 2D LGF.  We apply a smooth spherical cut-off function $f_\text{cut}(k/\kmax)$
\begin{equation}
f_\text{cut}(x) = \left\{\begin{array}{ll}
1 & : 0 \le x < \alpha,\\
3\left(\frac{1-x}{1-\alpha}\right)^2 - 2\left(\frac{1-x}{1-\alpha}\right)^3 & : \alpha \le x < 1,\\
0 & : 1 \le x\\
\end{array}\right.
\end{equation}
where $\kmax$ is the radius of a sphere inscribed in the Brillouin zone, to the divergent terms so that they and their derivatives are zero at the Brillouin zone edge.  The LGF semicontinuum correction is defined as the term remaining after subtracting the divergent and discontinuous terms,
\begin{equation}
\begin{split}
 \widetilde{G}^\text{sc}_\mathrm{AA}(\kv) &= \GAA(\kv) - \left(-k^{-2}\GAA^{(-2)}(\ku) - i k^{-1}\GAA^{(-1)}(\ku) + \GAA^{(0)}(\ku)\right)f_\text{cut}(k/\kmax)\\
 \widetilde{G}^{\text{sc}\dagger}_\mathrm{OA}(\kv) = \widetilde{G}^\text{sc}_\mathrm{AO}(\kv) &= \GAO(\kv) - \left(-i k^{-1}\GAO^{(-1)}(\ku) + \GAO^{(0)}(\ku)\right)f_\text{cut}(k/\kmax)\\
 \widetilde{G}^\text{sc}_\mathrm{OO}(\kv) &= \GOO(\kv) - \left(\GOO^{(0)}(\ku)\right)f_\text{cut}(k/\kmax)
\end{split}
\label{eqn:semicont}
\end{equation}
where we treat the semicontinuum terms through numerical inversion and subtraction of the divergent and discontinuous terms from Eq.~\eqref{eqn:gpower}.  For small values  of $k$ ($k < \kmax/10$), numerical truncation error in the divergent terms dominates the calculation.  Instead of direct subtraction of the divergent and discontinuous terms at small $k$, we define the leading order terms of the quadrants, $\Xi(\kv)$, as
\begin{equation}
\Xi(\kv) = \left(
\begin{array}{c|c}
-k^{-2}\GAA^{(-2)} & -ik^{-1}\GAO^{(-1)}\\
\hline
-ik^{-1}\GOA^{(-1)} & \GOO^{(0)}\\
\end{array}\right).
\label{eqn:xidef}
\end{equation}
We then calculate the semicontinuum correction for small $k$ as
\begin{equation}
\widetilde{G}^\text{sc}(\kv) = \left[\left(\eye - \Xi^{-1}(\kv)\widetilde{D}(\kv)\right)^{-1} - \eye\right]\Xi^{-1}(\kv) + ik^{-1}\left(\begin{array}{c|c}
\GAA^{(-1)} & 0\\
\hline
0 & 0\\
\end{array}\right) - \left(\begin{array}{c|c}
\GAA^{(0)} & \GAO^{(0)}\\
\hline
\GOA^{(0)} & 0\\
\end{array}\right).
\label{eqn:semicont_smallk}
\end{equation}

In 3D, a spherical harmonic expansion represents the angular variation in the Green function power series.  The spherical harmonic coefficients are
\begin{equation}
 \widetilde{G}^{(n)}_{\sigma\sigma'}(\ku) = \sum_{l=0}^{\Lmax}\sum_{m=-l}^{l}\widetilde{G}^{(n)}_{\sigma\sigma';lm}Y_{lm}(\ku)
 \label{eqn:3dgangle}
\end{equation}
where the series is truncated for $l > \Lmax$, and $\Lmax$ is chosen such that the $lm$ components above $\Lmax$ are less than $10^{-12}$ of the largest $lm$ component below $\Lmax$, and $Y_{lm}(\ku)$ are the normalized real spherical harmonics for $\ku = (\sin(\theta)\cos(\phi),\sin(\theta)\sin(\phi),\cos(\theta))$.  Due to inversion symmetry reciprocal space, for $n$ even, only the even $l$ terms are nonzero, and for $n$ odd, only the odd $l$ terms are nonzero.  The spherical harmonic components are found numerically by integrating over a uniform grid in $\phi$ and with Gaussian quadrature in $\theta$\cite{TrinkleLGF2008}.  Rotating back with the eigenvectors $e^{\mu}_{j\beta}$, and taking the inverse Fourier transform of the divergent and discontinuous terms gives the real space form\cite{TrinkleLGF2008}
\begin{equation}
 \underline{G}_{i\alpha,j\beta}^{(n)}(\Rv) = \sum_{l=0}^{\Lmax}\sum_{m=-l}^{l}\widetilde{G}^{(n)}_{i\alpha,j\beta;lm}Y_{lm}\left(\frac{\Rvuab{i}{j}}{|\Rvuab{i}{j}|}\right)(-1)^{l/2}\dfrac{i^{n}V}{2\pi^2}\int_0^{\kmax} dk\;k^{2+n}f_\text{cut}(k/\kmax)j_l(k|\Rvuab{i}{j}|),
\label{eqn:3dfourier}
\end{equation}
where $j_l(x)$ is the $l^\text{th}$ spherical Bessel function of the first kind, and $V$ is the volume of the unit cell.  The radial integrals are calculated numerically using adaptive Gauss-Kronrod integration.  Finally, the semicontinuum term is inverse Fourier transformed using a uniform k-point mesh to a tolerance of $10^{-7}$.  The error in the numerical inverse Fourier transform scales as $N_{k}^{4/d}$ for dimensionality $d$\cite{Ghazisaeidi2009}.

For a 2D LGF, such as one used to relax an infinite, straight line defect, we expand in a Fourier series.  The plane of the 2D LGF is normal to the threading vector $\vec{t}$ of the defect; $\vec{t}$ defines the periodicity of the defect.  By summing along $\vec{t}$, the Fourier transform of the LGF reduces to 2D in the Brillouin zone normal to $\vec{t}$. The Fourier coefficients are
\begin{equation}
 \widetilde{G}^{(n)}_{\sigma\sigma'}(\ku) = \sum_{m=0}^{\Mmax}\widetilde{G}^{(n)}_{\sigma\sigma';m}e^{i m\theta}
 \label{eqn:2dgangle}
\end{equation}
where $\ku = (\cos(\theta), \sin(\theta))$, and the Fourier series is truncated at $\Mmax$ so that components above $\Mmax$ are less than $10^{-12}$ of the largest component below $\Mmax$.  Due to inversion symmetry in $k$ space, the coefficients for even (odd) $n$ are only nonzero for even (odd) $m$.  Rotating back with the eigenvectors $e^{\mu}_{j\beta}$, and taking the inverse Fourier transform of the divergent and discontinuous terms of the 2D LGF gives the real space form\cite{TrinkleLGF2008}
\begin{equation}
 \underline{G}_{i\alpha,j\beta}^{(n)}(\Rv) = \sum_{m=0}^{\Mmax}\widetilde{G}^{(n)}_{i\alpha,j\beta;m}e^{im\phi_{\Rvuab{i}{j}}}(-1)^{m/2}\dfrac{i^{n}A}{2\pi}\int_0^{\kmax} dk\;k^{1+n}f_\text{cut}(k/\kmax)J_m(k|\Rvuab{i}{j}|),
\label{eqn:2dfourier}
\end{equation}
where $J_m(x)$ is the $m^\text{th}$ Bessel function of the first kind, and $A=V/|\vec{t}|$ is the area of the 2D unit cell.  We then evaluate these Bessel integrals numerically using adaptive Gauss-Kronrod integration.  Finally, the semicontinuum term is inverse Fourier transformed using a uniform k-point mesh to a tolerance of $10^{-7}$.

In summary, we: (1) Determine the acoustic/optical basis by calculating the eigenvectors $e^{\mu}_{j\beta}$ of the dynamical matrix at $\kv = 0$.  (2) Fourier transform the force-constant matrix and rotate the dynamical matrix into the acoustic/optical basis (Eq.~\eqref{eqn:drotsum}).  (3) Determine the power series coefficients of the dynamical matrix $\widetilde{D}^{(n)}_{\sigma\sigma',\mu\nu}(\ku)$ (Eq.~\eqref{eqn:dmatpowercoeff}) on an angular grid for analytic block inversion of the divergent and discontinuous terms of the LGF $\widetilde{G}^{(n)}_{\sigma\sigma',\mu\nu}(\ku)$ (Eq.~\ref{eqn:gpower} and Eq.~\ref{eqn:glambdas}).  (4) Determine the 2D Fourier coefficients $\widetilde{G}^{(n)}_{\sigma\sigma',m}$ (Eq.~\eqref{eqn:2dgangle}) or 3D spherical harmonic coefficients $\widetilde{G}^{(n)}_{\sigma\sigma',lm}$ (Eq.~\eqref{eqn:3dgangle}).  (5) Calculate the semi-continuum term $\widetilde{G}^\text{sc}(\kv)$ on a regular grid in reciprocal space by interpolating the Fourier or spherical harmonic expansion onto the grid and subtracting from the block inverse of the dynamical matrix (Eq.~\eqref{eqn:semicont} and Eq.~\eqref{eqn:semicont_smallk}).  (6) Inverse Fourier transform the divergent and discontinuity terms into real space analytically (Eq.~\eqref{eqn:2dfourier} or \eqref{eqn:3dfourier}), numerically inverse Fourier transform the semi-continuum terms.  (7) Add all contributions and rotate back to the original atomic basis to get the LGF in real space.  This automated algorithm directly calculates the lattice Green function from the force constant matrix with controllable numerical errors for a general crystal with any number of atoms in the crystal basis.  The numerical error scales with the number of k-points $N_{k}$ as $N_{k}^{4/d}$, where $d$ is the dimension of the system\cite{Ghazisaeidi2009}.  In the next section, we determine leading order corrections to the lattice Green function in terms of force-constant symmetry breaking.

\section{Square lattice model}
\label{sec:square}
\subsection{Single atom unit cell}
\begin{figure}
\centering
\includegraphics[width=1.5in]{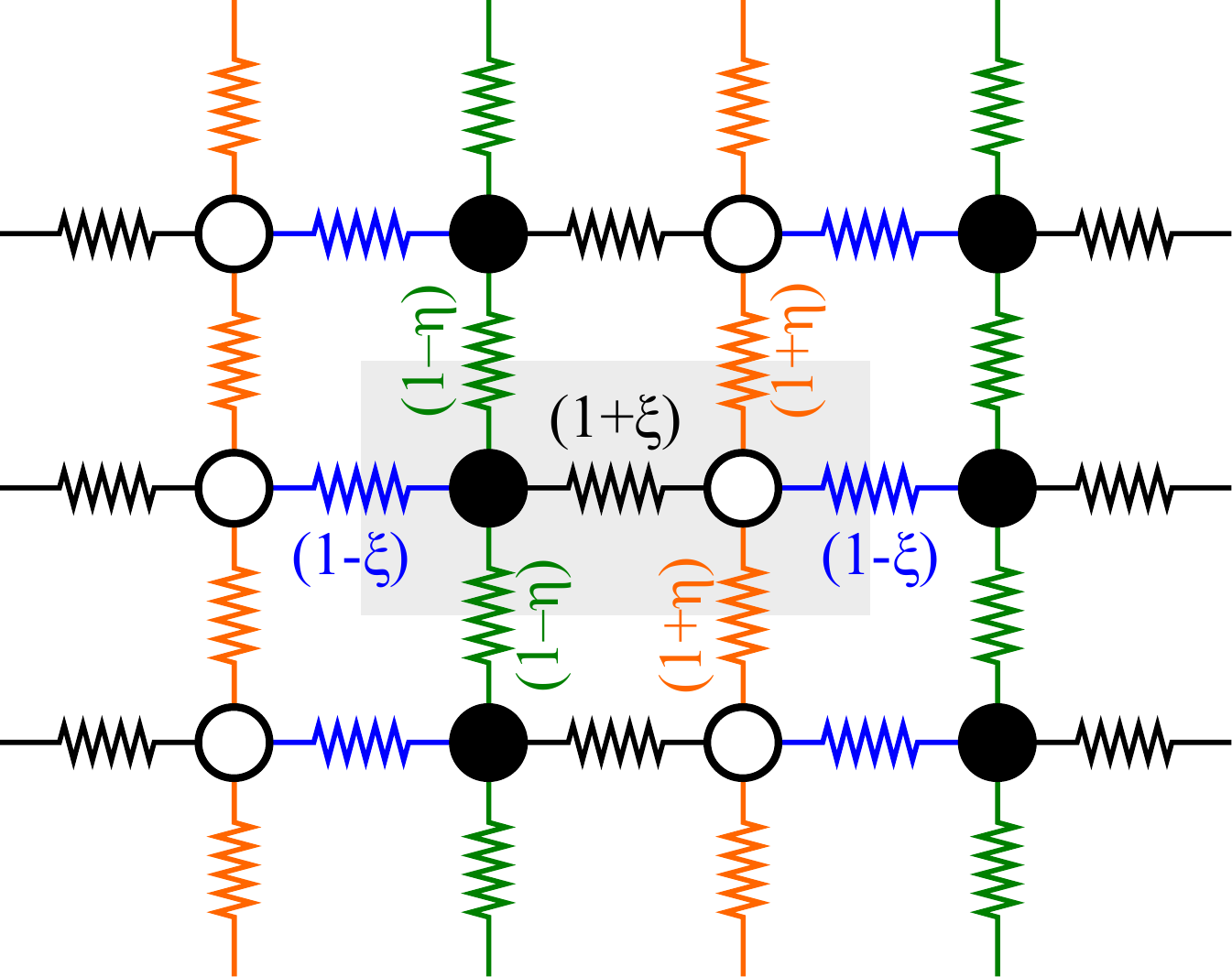}
\includegraphics[width=1.5in]{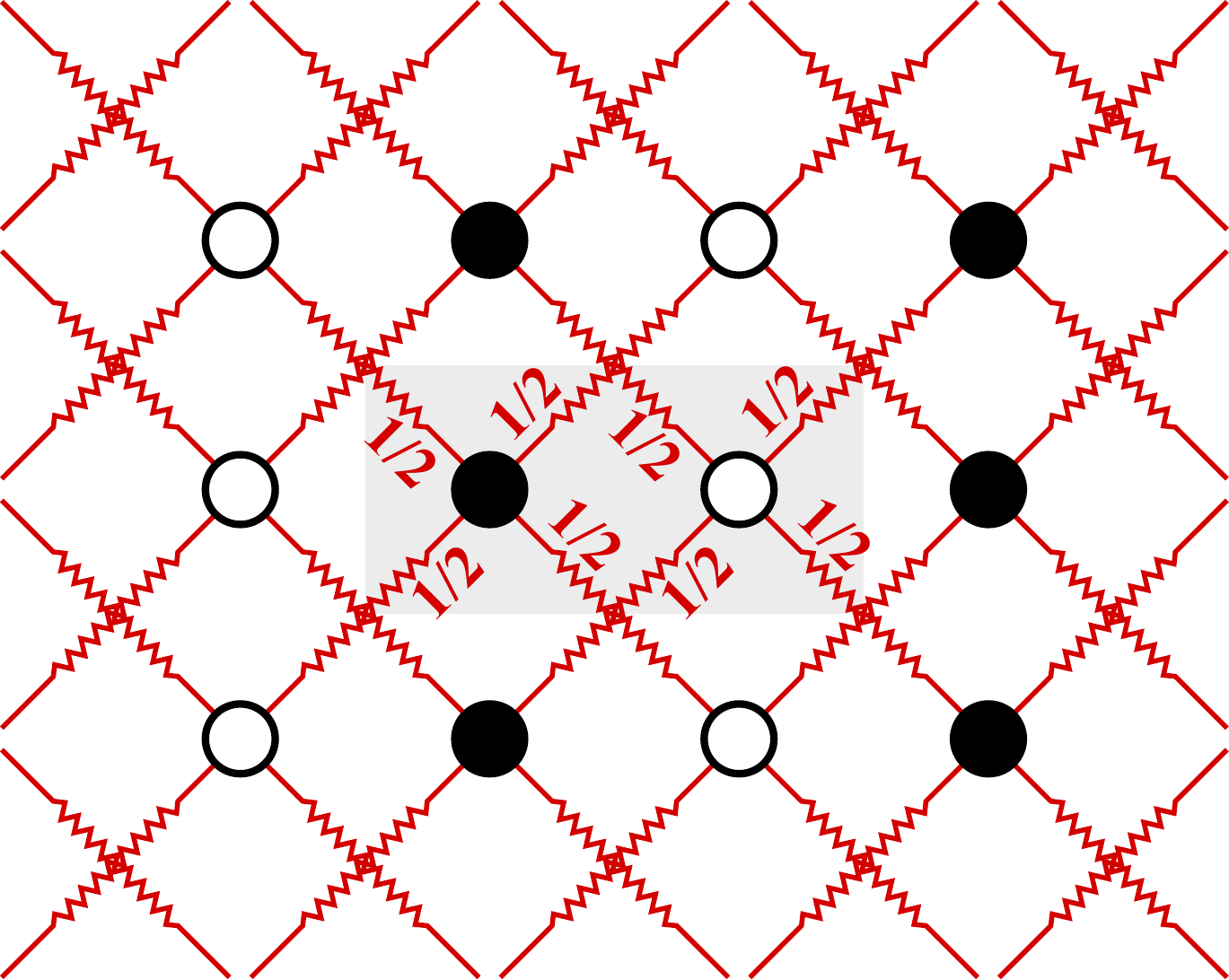}
\caption{Schematic of the two atom basis square lattice nearest neighbor interaction model with the spring constant perturbations $\xi$ and $\eta$ labeled.  The unit cell is shaded in gray.  Two different asymmetries are considered: $\eta$ describes the asymmetry between the white-white and black-black atom interactions (along $y$), and $\xi$ describes the asymmetry between the white-black unit cell spring and the white-black neighboring cell interactions (along $x$).  The diagonal second neighbor ``bond-bending'' springs of strength $1/2$ are shown in red; these springs stabilize the lattice and produce isotropic elastic response when $\eta=\xi=0$.}
\label{fig:springs}
\end{figure}

\begin{figure}
$\left(\begin{array}{c}\includegraphics[width=3.0in]{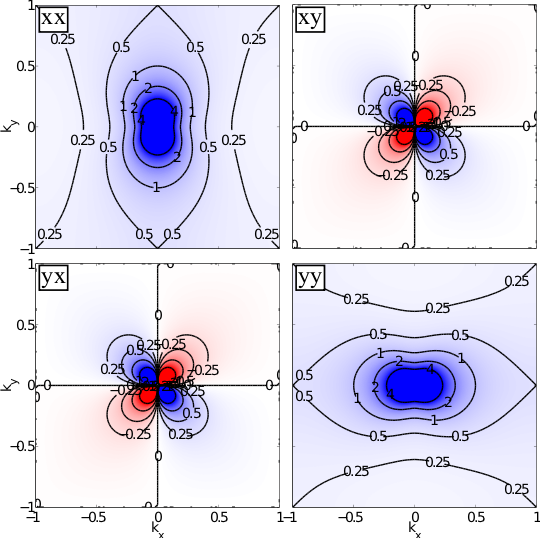}\end{array}\right)$
\caption{Contour plot of the Fourier transform of the lattice Green function for an isotropic square lattice ($\xi = \eta = 0$) over the first Brillouin zone (BZ).  The LGF is Hermitian, thus the bottom-left is the Hermitian conjugate of the upper-right.  All components show a second-order pole at $k=0$; c.f. Fig.~\protect\ref{fig:doublelgfpieces}.}
\label{fig:singleatom}
\end{figure}

Fig.~\ref{fig:springs} is a schematic of the doubled, elastically isotropic square lattice with first- and second-nearest-neighbor radial springs of spring constants 1 and 1/2 ($\xi=\eta=0$) that illustrates the additional terms added by a multiple-atom basis.  The single atom lattice with lattice vectors $\vec{a}_1 = (a, 0)$ and $\vec{a}_2 = (0, a)$ has elastic constants
\begin{equation}
 C_{11} = C_{22} = \frac32, \hspace{3em} C_{12} = C_{21} = \frac12, \hspace{3em} C_{66} = \frac12.
\end{equation}
This material is elastically isotropic ($2C_{66}=C_{11}-C_{12}$) with radial interactions for Cauchy symmetry ($C_{12} = C_{66}$); or 2D Poisson ratio of 1/3, Young's modulus of 4/3, and shear modulus of 1/2.  The Fourier transform of the LGF for the single atom case is shown in Fig.~\ref{fig:singleatom}. The divergent and discontinuous terms in the LGF for the single atom square lattice are
\addtolength{\arraycolsep}{0.1in}
\begin{equation}
\begin{split}
\widetilde{G}^{(-2)}(\kv) &= \frac{2}{3k^2a^2}
  \begin{pmatrix}
    2-\cos{2\theta} & -\sin{2\theta}\\
    -\sin{2\theta}  & 2+\cos{2\theta}
  \end{pmatrix}\\
\widetilde{G}^{(0)}(\kv) &= \frac{1}{72}
  \begin{pmatrix}
    7-4\cos{2\theta}+\cos{4\theta} & 0\\
    0                              & 7+4\cos{2\theta}+\cos{4\theta}
  \end{pmatrix}
\end{split}
\label{eqn:singleG20}
\end{equation}
\addtolength{\arraycolsep}{-0.1in}

\subsection{Doubled unit cell}
\begin{figure}
$\left(\begin{array}{c}\includegraphics[width=3.0in]{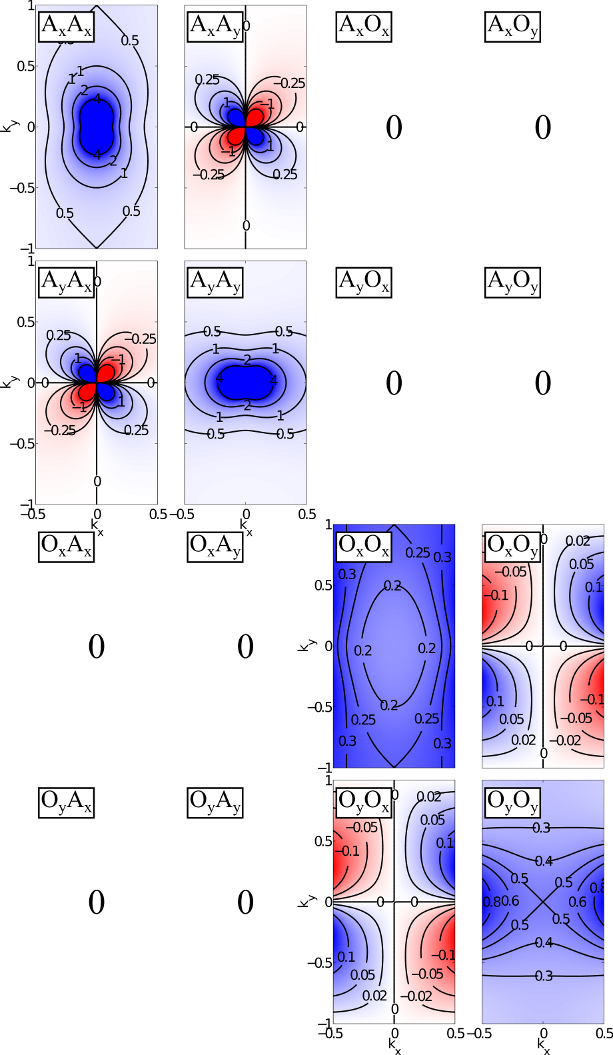}\end{array}\right)$
\caption{Contour plot of the Fourier transform of the lattice Green function for a doubled isotropic square lattice ($\xi = \eta = 0$) over the first Brillouin zone (BZ) in the acoustic(A)/optical(O) rotated basis.  The BZ is cut in half along the $k_x$ direction due to the doubling of the lattice along the $x$ direction.  Doubling the cell also results in the appearance of optical quadrants in the LGF.  The LGF is Hermitian, thus the bottom-left triangle is the Hermitian conjugate of the upper-right triangle.  The acoustic-acoustic quadrant, which corresponds to the collective motion of all the atoms, corresponds to the single atom unit cell LGF, and has a second-order pole at $k=0$; c.f. Fig.~\protect\ref{fig:doublelgfpieces}.}
\label{fig:doubleatom}
\end{figure}

Fig.~\ref{fig:springs} shows the doubled square lattice with lattice vectors $\vec{a}_1 = (2a, 0)$ and $\vec{a}_2 = (0, a)$
and crystal basis: $\vec{x}_b = (0,0)$ and $\vec{x}_w = (a,0)$ (for ``black'' and ``white'' atoms).  We consider the same interactions as in the single atom case.  The Fourier transform of the two atom LGF is shown in Fig.~\ref{fig:doubleatom}.  The acoustic-acoustic quadrant, which corresponds to the summed interactions of the black and white atoms, is the same as the LGF for the single atom case except for the halved Brillouin zone.  The divergent and discontinuous terms in the LGF are
\addtolength{\arraycolsep}{0.1in}
\begin{equation}
\begin{split}
 \widetilde{G}^{(-2)}_\text{AA} &= \frac{2}{3k^2a^2}
  \begin{pmatrix}
    2-\cos{2\theta} & -\sin{2\theta}\\
    -\sin{2\theta}  & 2+\cos{2\theta}
  \end{pmatrix}\\
 \widetilde{G}^{(-1)}_\text{OA} &= \widetilde{G}^{(-1)\dagger}_\text{AO} = 0\\
 \widetilde{G}^{(0)}_\text{AA} &= \frac{1}{72}
  \begin{pmatrix}
    7-4\cos{2\theta}+\cos{4\theta}   & 0\\
    0                                & 7+4\cos{2\theta}+\cos{4\theta}
  \end{pmatrix}\\
 \widetilde{G}^{(0)}_\text{OO} &= \frac{1}{6}
  \begin{pmatrix}
    1 & 0\\
    0 & 3
  \end{pmatrix}
\end{split}  
\label{eqn:doubleGexpand}
\end{equation}\\
\addtolength{\arraycolsep}{-0.1in}
with all other quadrants zero, and the same elastic constants as the single atom case.  

\begin{figure}
\centering
$\begin{array}{r}\dfrac{1}{(ka)^{2}}~\times~\left(
\begin{array}{c}\includegraphics[width=2.5in]{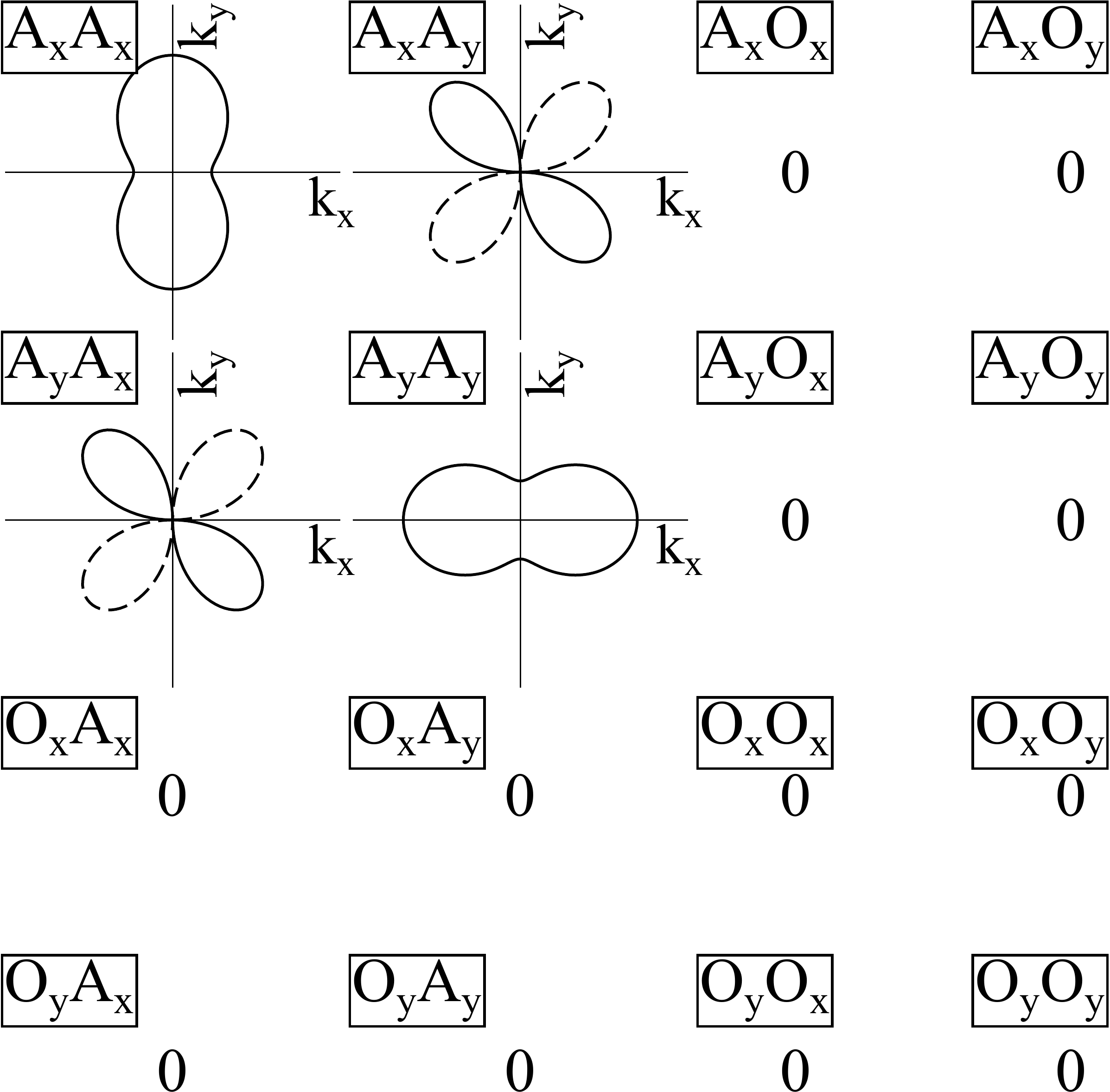}\end{array}
\right)\\
+~(ka)^{0}~\times~\left(
\begin{array}{c}\includegraphics[width=2.5in]{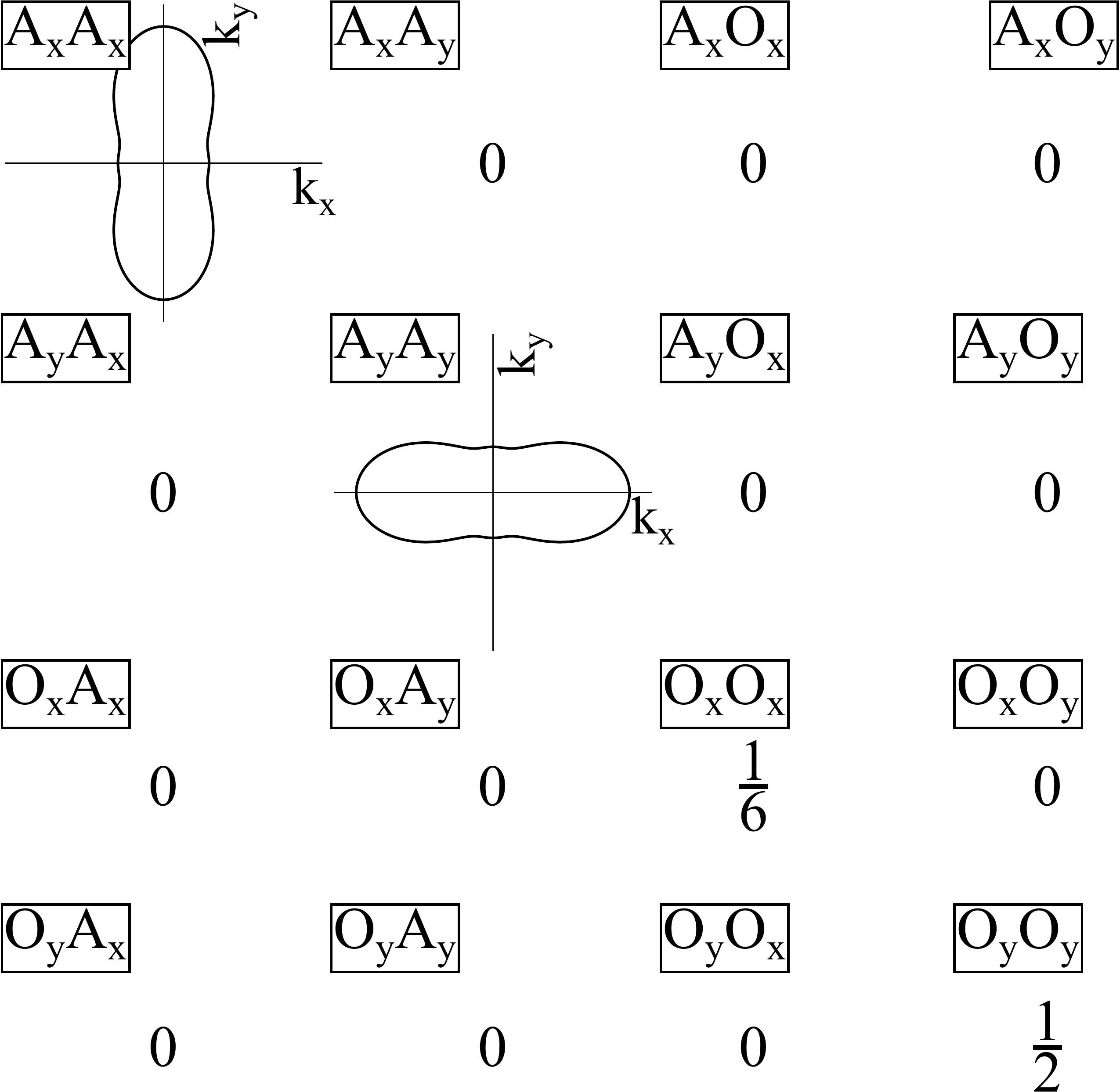}\end{array}
\right)\end{array}$
\caption{The directional dependence of the leading order divergent and discontinuous terms of the lattice Green function for a doubled square lattice.  The elastic Green function corresponds to the $1/k^2$ term.  The optical-optical quadrant is not discontinuous because the leading order corresponds to the inverse of the optical modes at $k=0$.  The magnitude along $\hat{k}$ in the polar plots is the multiplier for the LGF for direction $\hat{k}$; dashed lines correspond to negative multipliers.}
\label{fig:doublelgfpieces}
\end{figure}

Fig.~\ref{fig:doublelgfpieces} shows polar plots of the divergent and discontinuous LGF terms.  Inversion symmetry requires the acoustic-optical quadrant to be purely imaginary; since this is a doubled single atom system, the acoustic-optical quadrants are zero.  The acoustic-optical quadrants correspond to the response of the internal degrees of freedom of the system to elastic strain.  The doubled system behaves just as the single atom system; thus, there is no internal relaxation. The second-order pole and the discontinuity correction in the acoustic-acoustic quadrant are the same as the single atom case. This is because both cases describe the long range elastic behavior summed over all atoms.  The leading order optical-optical constants correspond to the inverse of the optical phonon frequencies at the $\Gamma$ point and are not discontinuous.  Since the cell is doubled along $\hat{x}$, the doubled system is stiffer along the $xx$ mode than the $yy$ mode because the $xx$ mode involves both bond-stretching and -bending springs, but the $yy$ mode only involves bond-bending springs.

\subsection{Breaking single atom symmetry}
\begin{figure}
\centering
$\begin{array}{r}\dfrac{i\xi}{ka}~\times~\left(
\begin{array}{c}\includegraphics[width=2.5in]{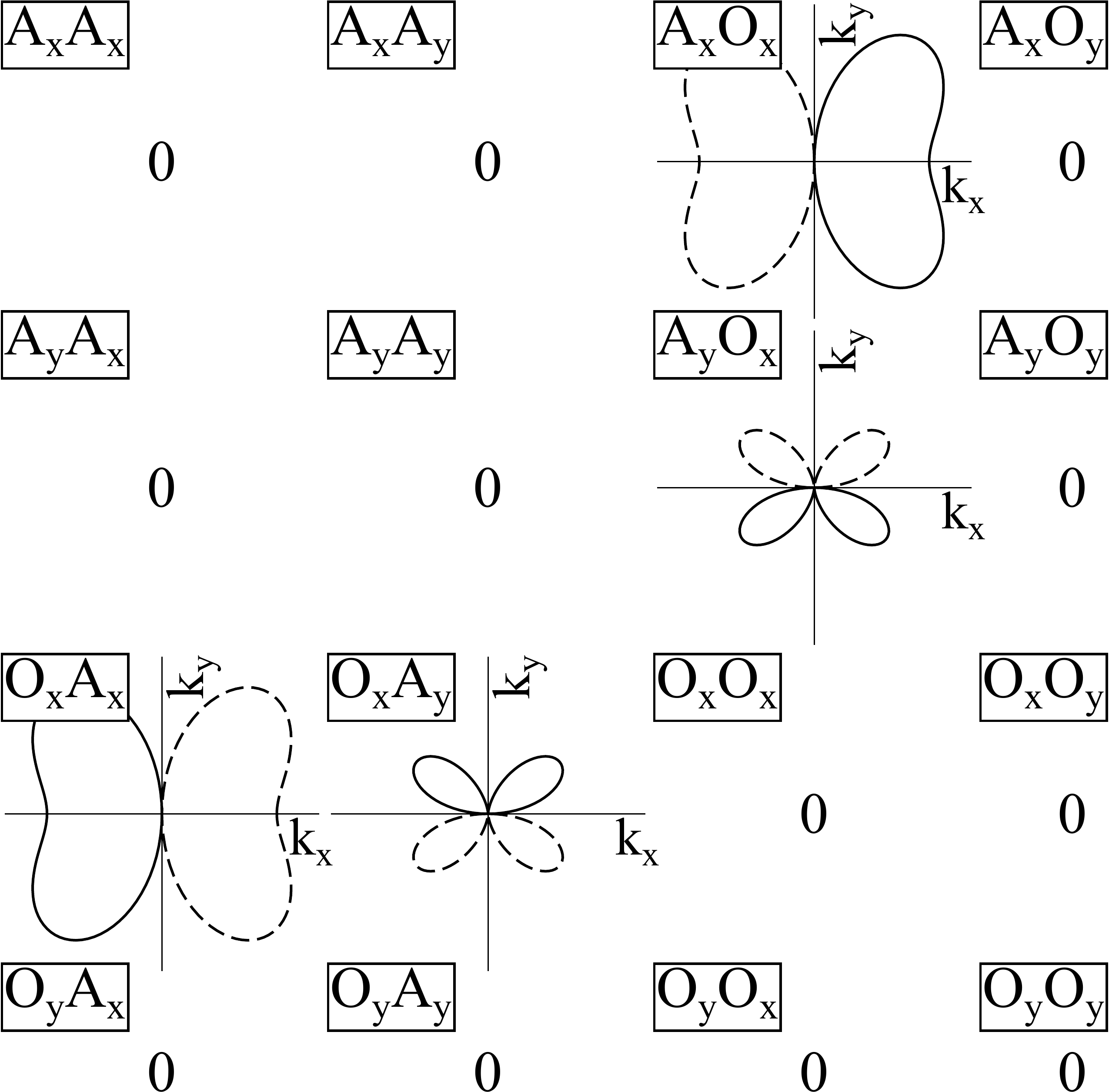}\end{array}
\right)\\
+~\eta (ka)^{0}~\times~\left(
\begin{array}{c}\includegraphics[width=2.5in]{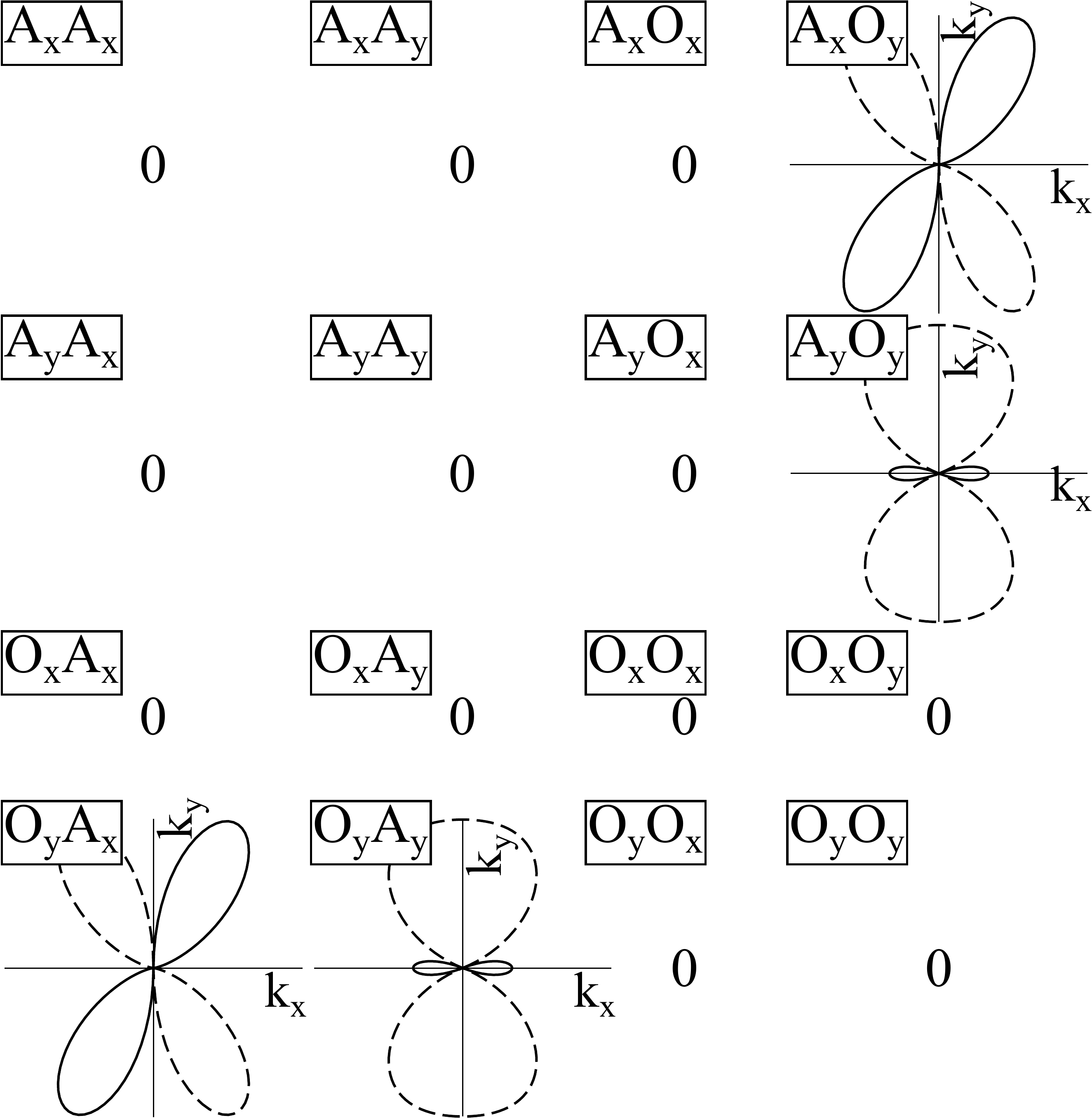}\end{array}
\right)\end{array}$
\caption{Leading order divergent term corrections of the lattice Green function for a doubled square
lattice as a function of the scaled relative spring constants $\eta$ (black-black/white-white coupling) and $\xi$ (black-white/white-black coupling). The magnitude along $\hat{k}$ in the polar plots is the multiplier for the LGF for direction $\hat{k}$; dashed lines correspond to negative multipliers.}
\label{fig:perturblgfpieces}
\end{figure}

We introduce small perturbations which break the single-atom symmetry of the doubled lattice to produce changes in the LGF.  We break the symmetry of the black/white atoms by changing the black-black spring constant to $(1+\eta)$ and the white-white spring
constant to $(1-\eta)$ as in Fig.~\ref{fig:springs}.  This has no effect on long range elastic behavior, and so the poles of the LGF are unmodified
by this perturbation.  To leading order in $\eta$, the interaction adds a discontinuity correction in the acoustic-optical mode,
\addtolength{\arraycolsep}{0.1in}
\begin{equation}
 \widetilde{G}^{(0)}_\text{OA} = \widetilde{G}^{(0)\textrm{T}}_\text{AO} = \frac{\eta}{12}
  \begin{pmatrix}
    0                            & 0\\
    2\sin{2\theta}-\sin{4\theta} & -3+4\cos{2\theta}+\cos{4\theta}\\
  \end{pmatrix} + O(\eta^2).
\end{equation}
\addtolength{\arraycolsep}{-0.1in}
The acoustic-acoustic discontinuity term is modified at second order in $\eta$.  Fig.~\ref{fig:perturblgfpieces} shows polar plots of these perturbations.  The acoustic-optical quadrants of the LGF correspond to the response of internal modes to collective motion.  However, breaking the black/white atom symmetry alone is not enough to generate internal relaxation in response to long range elastic waves as there is no adjustment to the $1/k^2$ or $i/k$ poles of the LGF.  The leading order adjustment occurs in the discontinuity correction which only depends upon the direction of $\hat{k}$.  Thus, there is an angular-dependent change due to the different stiffnesses of the black-black and white-white interactions.

To break symmetry and introduce internal relaxation, we set the in-cell black-white spring constant to $(1+\xi)$ and the black-white spring constant to neighboring cells to $(1-\xi)$ as shown in Fig.~\ref{fig:springs}.  This spring constant change causes a change in the long range elastic behavior of the model.  The $C_{11}$ elastic constant becomes, to second order, $3/2 - \xi^2/12$ due to internal relaxation of the atoms in the unit cell in response to strain, while $C_{22}$ remains $3/2$.  Linear order in $\xi$ introduces the acoustic-optical $i/k$ pole
\addtolength{\arraycolsep}{0.1in}
\begin{equation}
 \widetilde{G}^{(-1)}_\text{OA} = \widetilde{G}^{(-1)\dagger}_\text{AO} = \frac{2\xi i\cos{\theta}}{9ak}
  \begin{pmatrix}
    -2+\cos{2\theta} & \sin{2\theta}\\
     0               & 0
  \end{pmatrix} + O(\xi^2).
\end{equation}
\addtolength{\arraycolsep}{-0.1in}
At second order in $\xi$, there are changes in the acoustic-acoustic second-order pole and discontinuity correction, and the optical-optical discontinuity correction.  Fig.~\ref{fig:perturblgfpieces} shows polar plots of the first-order perturbations.  By breaking the in-cell symmetry, an imaginary term now appears in the acoustic-optical quadrants of the LGF.  This $i/k$ pole comes from internal relaxation of atoms in the unit cell due to a long-wavelength elastic wave.  Breaking the internal symmetry causes the black and white atoms to respond differently to a long range strain and shift from their strained simple square lattice sites in response.  This term is important for describing the internal response of a multi-atom basis crystal and does not appear in the single atom Bravais lattice case.

\subsection{The simple square lattice in real space}
\begin{figure}
$\left(\begin{array}{c}\includegraphics[width=3.0in]{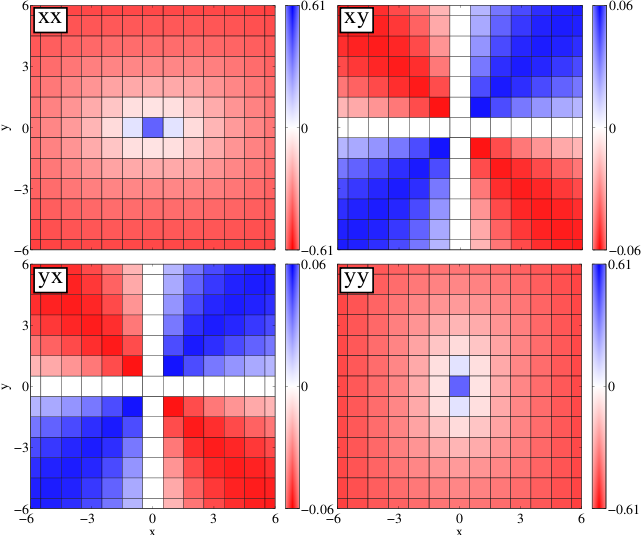}\end{array}\right)$
\caption{LGF in real space for an isotropic square lattice ($\xi = \eta = 0$).  The LGF is only defined at crystal sites.  Red and blue coloring correspond to negative and positive values of the LGF at those crystal sites.  The two atom LGF is identical to the single atom LGF at the crystal sites to $10^{-7}$ for a $80 \times 80$ ($40 \times 80$) k point mesh for the single (double) atom cell.}
\label{fig:gmat1a}
\end{figure}

Fig.~\ref{fig:gmat1a} shows the real space LGF for the single and doubled unit cells in crystal space by taking the inverse Fourier transform.  Without interaction perturbations, both are identical to $10^{-7}$ for a $40\times80$ k-point mesh for the doubled cell and a $80\times80$ k-point mesh for the single cell.  Since both have the same interactions, both methods yield the same harmonic response in a different basis.  The LGF plot shows the linear relationship between force and displacement for atoms separated by $\Rvuab{i}{j}$.  The elastic LGF is logarithmic in $|\Rvuab{i}{j}|$ and dominated by the inverse transform of the $k^{-2}$ pole at long range.

\begin{figure}
$\begin{array}{r}\xi\left(
\begin{array}{c}\includegraphics[width=3.0in]{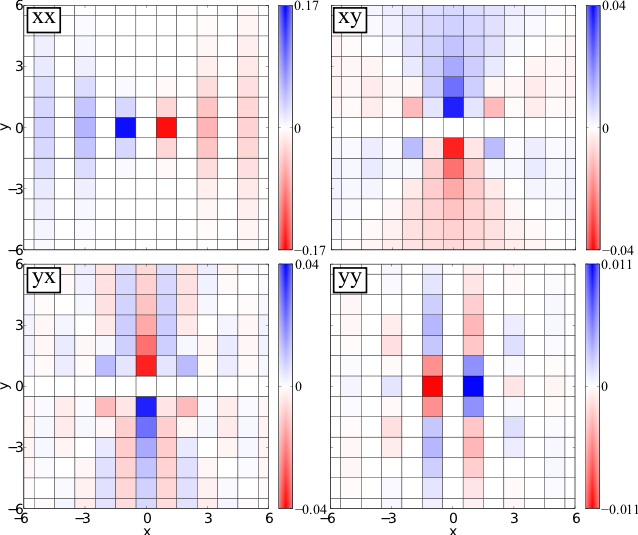}\end{array}
\right)\\
+\eta\left(
\begin{array}{c}\includegraphics[width=3.0in]{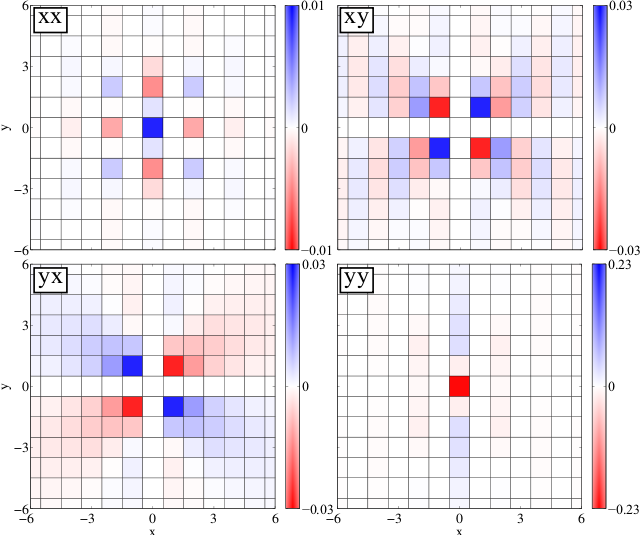}\end{array}
\right)\end{array}$
\caption{Change to the black atom LGF due to the $\xi$ (black-white/white-black coupling) and $\eta$ (black-black/white-white coupling) perturbations.  The LGF is only defined at crystal sites, with red and blue coloring corresponding to negative and positive values.  The change in LGF for white atoms is opposite to the change for black atoms.}
\label{fig:gmatxieta}
\end{figure}

To first order, neither the $\eta$ or $\xi$ perturbations modify the $k^{-2}$ pole of the LGF.  Therefore, at long range in real space, the LGF is still dominated by the original isotropic elastic behavior.  Considering the inverse transforms of the perturbed interactions, the $\eta$ perturbation results in opposite $yy$ response for the black/black and white/white interactions as expected since it is a change in the $yy$ spring strengths.  The $\xi$ perturbation results in opposite $xx$ response for the black/white and white/black interactions.  The effects of $\xi$ are seen in the $i k^{-1}$ pole, and only visible at short to intermediate range with a decay of $1/R$, while $\eta$ first appears in $k^{0}$ component of the LGF, and is only seen at short range since it decays as $1/R^2$. Figure~\ref{fig:gmatxieta} shows the first-order real-space response to $\xi$ and $\eta$ perturbations.

\section{Conclusion}
The direct, automated algorithm can efficiently and accurately calculate the lattice Green function for general crystals with more than one atom in the unit cell basis and arbitrary interactions.  Additional terms describing the response of internal degrees of freedom of the system corresponding to optical modes appear in this formalism that did not appear in a treatment for a Bravais lattice.  Including the additional optical terms of the lattice Green function extendes the previous automated calculation of the LGF for long-range interactions\cite{TrinkleLGF2008} to general crystals.  This technique  efficiently calculates defect structures in HCP metals such as Mg\cite{Yasi2009}, Ti\cite{Ghazisaeidi2012}, as well as semiconductors and intermetallics using flexible boundary condition methods\cite{Woodward2002}.  In particular, reducing the number of atoms required for accurate calculation of isolated dislocation core geometries provides efficient use of density functional theory.

\begin{acknowledgments}
This research was sponsored by NSF through the GOALI program, grant 0825961, and with the support of General Motors, LLC.  The code is publicly available due to hosting support of the NSF MatForge project at \url{http://matforge.org/redmine/projects/lgf}.
\end{acknowledgments}

\end{document}